\documentstyle[12pt]{article}
\begin{document}

\def\ds{\displaystyle}
\def\beq{\begin{equation}}
\def\eeq{\end{equation}}
\def\bea{\begin{eqnarray}}
\def\eea{\end{eqnarray}}
\def\beeq{\begin{eqnarray}}
\def\eeeq{\end{eqnarray}}
\def\ve{\vert}
\def\vel{\left|}
\def\brll{B\rar\rho \ell^+ \ell^-}
\def\ver{\right|}
\def\nnb{\nonumber}
\def\ga{\left(}
\def\dr{\right)}
\def\aga{\left\{}
\def\adr{\right\}}
\def\lla{\left<}
\def\rra{\right>}
\def\rar{\rightarrow}
\def\nnb{\nonumber}
\def\la{\langle}
\def\ra{\rangle}
\def\ba{\begin{array}}
\def\ea{\end{array}}
\def\tr{\mbox{Tr}}
\def\ssp{{\Sigma^{*+}}}
\def\sso{{\Sigma^{*0}}}
\def\ssm{{\Sigma^{*-}}}
\def\xis0{{\Xi^{*0}}}
\def\xism{{\Xi^{*-}}}
\def\qs{\la \bar s s \ra}
\def\qu{\la \bar u u \ra}
\def\qd{\la \bar d d \ra}
\def\qq{\la \bar q q \ra}
\def\gGgG{\la g^2 G^2 \ra}
\def\q{\gamma_5 \not\!q}
\def\x{\gamma_5 \not\!x}
\def\g5{\gamma_5}
\def\sb{S_Q^{cf}}
\def\sd{S_d^{be}}
\def\su{S_u^{ad}}
\def\ss{S_s^{??}}
\def\ll{\Lambda}
\def\lb{\Lambda_b}
\def\sbp{{S}_Q^{'cf}}
\def\sdp{{S}_d^{'be}}
\def\sup{{S}_u^{'ad}}
\def\ssp{{S}_s^{'??}}
\def\sig{\sigma_{\mu \nu} \gamma_5 p^\mu q^\nu}
\def\fo{f_0(\frac{s_0}{M^2})}
\def\ffi{f_1(\frac{s_0}{M^2})}
\def\fii{f_2(\frac{s_0}{M^2})}
\def\O{{\cal O}}
\def\sl{{\Sigma^0 \Lambda}}
\def\es{\!\!\! &=& \!\!\!}
\def\ar{&+& \!\!\!}
\def\ek{&-& \!\!\!}
\def\cp{&\times& \!\!\!}
\def\se{\!\!\! &\simeq& \!\!\!}
\def\hml{\hat{m}_{\ell}}
\def\rr{\hat{r}_{\Lambda}}
\def\ss{\hat{s}}


\renewcommand{\textfraction}{0.2}    
\renewcommand{\topfraction}{0.8}

\renewcommand{\bottomfraction}{0.4}
\renewcommand{\floatpagefraction}{0.8}
\newcommand\mysection{\setcounter{equation}{0}\section}

\def\baeq{\begin{appeq}}     \def\eaeq{\end{appeq}}
\def\baeeq{\begin{appeeq}}   \def\eaeeq{\end{appeeq}}
\newenvironment{appeq}{\beq}{\eeq}
\newenvironment{appeeq}{\beeq}{\eeeq}
\def\bAPP#1#2{
 \markright{APPENDIX #1}
 \addcontentsline{toc}{section}{Appendix #1: #2}
 \medskip
 \medskip
 \begin{center}      {\bf\LARGE Appendix #1 :}{\,\,\,\,\Large\bf #2}
\end{center}
 \renewcommand{\thesection}{#1.\arabic{section}}
\setcounter{equation}{0}
        \renewcommand{\thehran}{#1.\arabic{hran}}
\renewenvironment{appeq}
  {  \renewcommand{\theequation}{#1.\arabic{equation}}
     \beq
  }{\eeq}
\renewenvironment{appeeq}
  {  \renewcommand{\theequation}{#1.\arabic{equation}}
     \beeq
  }{\eeeq}
\nopagebreak \noindent}

\def\eAPP{\renewcommand{\thehran}{\thesection.\arabic{hran}}}

\renewcommand{\theequation}{\arabic{equation}}
\newcounter{hran}
\renewcommand{\thehran}{\thesection.\arabic{hran}}

\def\bmini{\setcounter{hran}{\value{equation}}
\refstepcounter{hran}\setcounter{equation}{0}
\renewcommand{\theequation}{\thehran\alph{equation}}\begin{eqnarray}}
\def\bminiG#1{\setcounter{hran}{\value{equation}}
\refstepcounter{hran}\setcounter{equation}{-1}
\renewcommand{\theequation}{\thehran\alph{equation}}
\refstepcounter{equation}\label{#1}\begin{eqnarray}}


\newskip\humongous \humongous=0pt plus 1000pt minus 1000pt
\def\caja{\mathsurround=0pt}


\title{
 {\small \begin{flushright}
\today
\end{flushright}}
       {\Large
                 {\bf
 Double-Lepton Polarization Asymmetries and Polarized Forward Backward Asymmetries
in The Rare $b \to s \ell^+ \ell^-$  Decays in a Single Universal
Extra Dimension Scenario
                 }
         }
      }
\author{\\
{\small V. Bashiry$^1$\thanks {e-mail: bashiry@ciu.edu.tr}, M.
Bayar$^2$\thanks{e-mail: mbayar@newton.physics.metu.edu.tr}, K.
Azizi$^2$\thanks {e-mail: e146342@metu.edu.tr}\,\,, }\\
{\small $^1$ Engineering Faculty, Cyprus International University,}
\\ {\small Via Mersin 10, Turkey }\\{\small $^2$
Physics Department, Middle East Technical University,}\\
{\small 06531 Ankara, Turkey}}
\date{}
\begin{titlepage}
\maketitle
\thispagestyle{empty}

\begin{abstract}
We study the double-lepton polarization asymmetries and single and
double-lepton polarization forward backward asymmetries of  the
rare $b \to s \ell^+ \ell^-$  mode within the Standard Model and
in the Appelquist-Cheng-Dobrescu model, which is a new physics
scenario with a single universal extra dimension. In particular,
we examine the sensitivity of the observables to the radius $R$ of
the compactified extra-dimension, which is the new parameter in
this new physics model.
\end{abstract}

~~~PACS numbers: 12.60.-i, 13.25.Hw
\end{titlepage}

\section{Introduction}

The standard model (SM) has been successful theory in reproducing
almost all
 experimental data about the interaction of gauge bosons and fermions.
 However, the SM is not regarded as a full theory, since it can not address
 some issues such as stability of the scalar sector under radiative corrections,
 gauge and fermion mass hierarchy, matter-antimatter
 asymmetry, number of generations and so on.  For these reasons, It is well known that the
 SM need to be extended in a way where the SM can be considered as a low energy manifestation
  of some fundamental theory or alternative theories must used instead
 of the SM.

Extra dimension~(ED) model with a flat metric proposed by Arkani
et. al.~\cite{Arkani} or with small compactification radius is one
of the candidates trying to shed light on some of those issues as
well as to provide a unified framework for gravity and the other
interactions together with  a connection with the string theory
\cite{Arkani}. It can be categorized
  in terms of the mechanism of new physics (NP) where the SM fields are
constrained to move in the usual three spatial dimensions ($D_3$
bran)
 or can propagate in the extra dimensions (the bulk). The last one
can be categorized as non--universal extra dimension (NUED)  and
universal extra dimension (UED). In the non universal model, the
gauge bosons propagate into the bulk,
 but the fermions are confined to $D_3$ bran. In contrast, the UED, which is the most democratic, allows
fields to propagate into the bulk. The UED can be considered as a generalization
 of the usual SM to a $D_{3+N}$ bran where N is the number of the extra dimensions\cite{Colider}. The first proposal for
using large (TeV) extra dimensions in the SM proposed by I.
Antoniadis~\cite{Antoniadis} who worked out the main consequences.
The model introduced by Appelquist, Cheng and Dobrescu
(ACD)~\cite{ACD} is the most simple example of the UED where just
single universal extra dimension  is considered. This model has
only one free parameter in addition to the SM parameters and that
is the compactification scale $R$. The ACD model is a particular
model within the same general idea proposed in~\cite{Antoniadis}.
Mass of the Kaluza-Klein(KK) particles are inversely proportional
to $R$, then, above some compactification scale $1/R$ the models
are higher dimensional theories whose equivalent description in
four dimensions includes the ordinary SM fields together with
towers of their Kaluza-Klein (KK) excitations and other fields
having no standard model partners\cite{Colangelo:2006gv}.

Two types of study can be conducted to explore extra dimensions. In the direct search, the
center of mass energy of colling particles must be increased to produce Kaluza-Klein(KK)
 excitation states, where KK excitation states
are supposed to produce in pair by KK number conservation.
On the other hand, we can investigate UED effects, indirectly. The indirect
search at tree-level, where KK excitations can contribute as a mediator, is suppressed by KK
number conservation. On the contrary, the same states can contribute to the quantum loop level where the KK number
conservation is broken. As a result, flavor changing neutral current(FCNC) transition induced by quantum loop level
can be considered as a good tool for studying KK effects. The collider signatures and phenomenology
of UED have been studied by Ref. \cite{Colider} and \cite{Buras1,Buras2}, respectively.
These studies have provided a theoretical framework to investigate some
inclusive and exclusive decays with the ACD.

FCNC and CP-violating are indeed the most sensitive probes of NP
contributions to penguin operators. Rare decays, induced by FCNC
of $b \rar s(d)$ transitions are at the forefront of our quest to
understand flavor and the origins of CP violation asymmetry (CPV),
offering one of the best probes for NP beyond the SM, in
particular to probe extra dimension. In this regard, the
semileptonic and pureleptonic B decays have been studied with UED
scenario\cite{Buras1}--\cite{Pakistan}. They have obtained that
the inclusive and exclusive semileptonic and pureleptonic decays
are sensitive to the new parameter coming  out of the one
universal extra dimensions i.e., compactification scale $1/R$.

 New physics
effects manifest themselves in rare decays in different ways: NP
can contribute through the new Wilson coefficients or the new
operator structure in the effective Hamiltonian, which is absent
in the SM. A crucial problem in the new physics search within
flavour physics in the exclusive decays is the optimal separation
of NP effects from uncertainties. It is well known that inclusive
decay modes are dominated by partonic contributions;
non--perturbative corrections are in general rather
smaller\cite{Hurth}. Also, ratios of exclusive decay modes such as
asymmetries for $B\rar K(~K^\ast,~\rho,~\gamma)~ \ell^+ \ell^-$
decays \cite{R4621}--\cite{R4622} are well studied for NP search.
Here, large parts of the hadronic uncertainties partially cancel
out. The universal extra dimension with only one UED belongs to
the classes of NP, where the Wilson coefficients is modified by KK
contributions \cite{Buras1, Buras2} in the penguin and box
diagrams.  Obviously, these modifications will affect the physical
observables. In this connection, we try to investigate the effects
of one universal extra dimension on the double--lepton
polarization and polarized forward--backward(FB) asymmetries of
the $b\to s \ell^+\ell^-$ transition, where the single-lepton
polarization asymmetries in the same scenario studied in
Ref.\cite{Colangelo:2006gv}.  Also, it is well known that the
study of the lepton polarization asymmetries are in particular
interesting since they are sensitive to  the structure of
interactions which can be used as a good tool to test not only the
SM but also its extensions\cite{R4621}.  Moreover, It is already
noted that the study of some of the single lepton polarization
asymmetries which are small  might not provide sufficient number
of observables for checking the structure of the effective
Hamiltonian. On the other hand, Considering the  polarizations of
both leptons, which are supposed to measure simultaneously, we are
able to establish the maximum number of independent polarization
observables\cite{Bensalem:2002ni}.

 The plan of the paper is the following:   In
Section 2 we recall the effective Hamiltonian inducing $b \to s
\ell^+ \ell^-$ transitions in the SM and in the ACD model, together
with the definition of the polarization asymmetries considering in
our study. In this Section, we also present the lepton polarization
and FB asymmetries for inclusive $b \to s \ell^+ \ell^-$ transition.
  Section 3  includes numerically analyzing the physical observables and
   conclusions are presented in section 4.
\section{Matrix element $b\rightarrow s\ell^{+}\ell^{-}$ in the ACD model}
In the SM,  the QCD corrected Hamiltonian for the transitions
$b\rightarrow s\ell^{+}\ell^{-}$ can be achieved by integrating out
the heavy quarks and the heavy electroweak bosons\cite{aali}:
 \begin{equation} H_W\,=\,4\,{G_F
\over \sqrt{2}} V_{tb} V_{ts}^\ast \sum_{i=1}^{10} C_i(\mu) O_i(\mu)
\label{hamil}
\end{equation}
\noindent
obtained by a  renormalization group evolution
  from the electroweak scale down to $\mu\simeq m_b$.
$G_F$ is the Fermi constant, $V_{ij}$ are elements of the
Cabibbo-Kobayashi-Maskawa (CKM) matrix, $O_i$ are the local
operators and
 $C_{i}$ are Wilson coefficients calculated
 in naive dimensional regularization (NDR) scheme
 at the leading order (LO), next-to-leading order (NLO),
 and next-to-next-to leading order (NNLO) in the SM\cite{R23}--\cite{NNLL}.

The effect of the new states predicted in the ACD model  comes
through the modification of the Wilson coefficients and the operator
structures remain the same as SM. In particular,  the coefficients
acquire a dependence on the  compactification radius $R$.
Considering the KK modes effects in the penguin and box diagrams,
the above coefficients have been obtained at
LO~\cite{Buras1,Buras2}. Clearly, they depend on the additional ACD
parameter i.e.,  $R$. For large values of $1/R$ the SM values of the
Wilson coefficients can be achieved. In general, the coefficients
can be expressed in terms of functions $F(x_t,1/R )$, with
$x_t=\displaystyle{ m_t^2 \over M_W^2}$ and $m_t$ is the top quark
mass. These functions are generalizations of the corresponding SM
functions such as $F_0(x_t)$ according to: \beq
F(x_t,1/R)=F_0(x_t)+\sum_{n=1}^\infty F_n(x_t,x_n) \,, \label{fxt}
\eeq
 with
$x_n=\displaystyle{ m_n^2 \over M_W^2}$ and $m_n=\displaystyle{n
\over R}$. The Glashow-Illiopoulos-Maiani (GIM) mechanism guarantees
the finiteness of the Eq. (\ref{fxt}) and fulfills the condition
$F(x_t,1/R) \to F_0(x_t)$ when $R \to 0$ \cite{Buras1,Buras2}.
However, as far as $1/R$ is taken in the order of a few hundreds of
GeV, the coefficients differ from the SM value: in particular,
$C_{10}$ is enhanced and $C_7$ is suppressed. Obviously, such
deviations could be seen in various observables in the inclusive and
exclusive B decays.

In the following, we only consider the contribution of the operators
$O_7$, $O_9$, and $O_{10}$. Note that we ignore the $O(\alpha_s)$
correction coming from one gluon exchange in the matrix element of
the operator ${\cal O}_9$ \cite{R5734},
 one--loop corrections to the
four--quark operators $O_1$--$O_6$ which are small \cite{R25} and
also the long-distance resonance effects. However, a more
complimentary and supplementary analysis of the above decay has to
be taken into account not only the long-distance contributions,
which have their origin in real intermediate $c\bar{c}$
family\cite{Deshpande} but also $O(\alpha_s)$ correction. The Wilson
coefficients $C_7$, $C_9$, and $C_{10}$ in the ACD are real and
their explicit expressions can be found in
Refs.\cite{Buras1,Buras2}.

In order to compute the polarization asymmetries, one has to choose
a reference frame to define the spin directions. A reference frame
can be chosen in the center of mass (CM) of the leptons where they
move back to back. In such reference frames, if we suppose that
$\ell^-$ moves in the $z$ positive direction and the fact that
momentum must conserve, the $s$ and $b$ quarks move in the same
direction. In this reference frame, the 4-vector $s^\mu_{\ell^-}$
can be obtained as follows after the Lorentz boost from its rest
frame\cite{Bensalem:2002ni}:

\beq s^\mu_{\ell^-} = \left\{ {P\over m_\ell} s^-_z , s^{-}_x ,
s^{-}_y, {\sqrt{P^2 + m_\ell^2} \over m_\ell} s^{-}_z \right\}
~~,~~~~ s^\mu_{\ell^+} = \left\{ -{P\over m_\ell} s^+_z , s^+_x ,
s^{+}_y, {\sqrt{P^2 + m_\ell^2} \over m_\ell} s^{+}_z \right\} ~.
\label{frame} \eeq

The $\hat{s}=\frac{q^{2}}{m_{b}^{2}}$ dependent double--lepton
polarization asymmetries ${\cal P}_{ij}$ are obtained by evaluating

\beq {\cal P}_{ij} =\frac{\Big[\frac{d\Gamma({\bf s^{+}}={\bf
\hat{i}},{\bf
      s^{-}}={\bf \hat{j}})}{d\hat{s}}-\frac{d\Gamma({\bf s^{+}}={\bf
      \hat{i}},{\bf s^{-}}={-\bf \hat{j}})}{d\hat{s}}\Big]
  -\Big[\frac{d\Gamma({\bf s^{+}}={-\bf \hat{i}},{\bf s^{-}}={\bf
      \hat{j}})}{d\hat{s}}-\frac{d\Gamma({\bf s^{+}}={-\bf
      \hat{i}},{\bf s^{-}}={-\bf \hat{j}})}{d\hat{s}}\Big]}
{\Big[\frac{d\Gamma({\bf s^{+}}={\bf \hat{i}},{\bf s^{-}}={\bf
      \hat{j}})}{d\hat{s}}+\frac{d\Gamma({\bf s^{+}}={\bf
      \hat{i}},{\bf s^{-}}={-\bf \hat{j}})}{d\hat{s}}\Big]
  +\Big[\frac{d\Gamma({\bf s^{+}}={-\bf \hat{i}},{\bf s^{-}}={\bf
      \hat{j}})}{d\hat{s}}+\frac{d\Gamma({\bf s^{+}}={-\bf
      \hat{i}},{\bf s^{-}}={-\bf \hat{j}})}{d\hat{s}}\Big]},
\eeq
where ${\hat i}$ and ${\hat j}$ are unit vectors\cite{Fukae}.

With our choice of reference frame in Eq.~(\ref{frame}), the decay
happens in two dimensions i.e., the $yz$ plane. In this frame, just
the components of the spin can be in the $\hat x$ direction.
Therefore, any terms including the spin along the $\hat x$ direction
are as a result of either the dot product of two spins or
triple-product correlation with one spin along the $\hat x$
direction(i.e.,\ ${\cal P}_{xx}$, ${\cal P}_{xy}$, and ${\cal
P}_{xz}$). This holds even in the presence of any extension of the
SM. Among these quantities, ${\cal P}_{xy}$ and ${\cal P}_{xz}$ are
attractive which probes the imaginary parts of the products of
Wilson coefficients\cite{Bensalem:2002ni}. The expressions for the
${\cal P}_{ij}$ asymmetries in $b\to s \ell^+ \ell^-$ can be derived
from the transition  amplitude
\bea {\cal M}={G_F \over \sqrt{2}} V_{tb} V_{ts}^* {\alpha \over
\pi} \, &\Big[&C_9(\mu, 1/R) \, {\bar s}_L \gamma_\mu b_L {\bar
\ell} \gamma_\mu \ell + C_{10}(\mu, 1/R) \, {\bar s}_L \gamma_\mu
b_L {\bar \ell} \gamma_\mu \gamma_5 \ell \nnb \\ &-& 2 C_7(\mu,
1/R)\, {q^\nu \over q^2} \, \left[ m_b
 {\bar s}_L i \sigma_{\mu \nu} b_R + m_s {\bar s}_R i \sigma_{\mu \nu} b_L \right]
 {\bar \ell}
\gamma_\mu \ell \Big].  \label{ampl} \eea Note that the measurements
of such asymmetries can give more information about
 the Wilson coefficients~\cite{R4621}.

The ${\cal P}_{ij}$ take the form
\begin{eqnarray} \label{2}
{\cal P}_{xx}&=&\frac{1}{\Delta} \Bigg
\{24Re[C_{7}(\mu, 1/R)C_{9}^\ast(\mu, 1/R)]\frac{\hat{m}{_{\ell}^{2}}}{\hat{s}}
+4|C_{7}(\mu, 1/R)|^{2}\frac{(-1+\hat{s})\hat{s}+2(2+\hat{s})\hat{m}{_{\ell}^{2}}}{\hat{s}^{2}}\nonumber
\\&&+(|C_{9}(\mu, 1/R)|^{2}-|C_{10}(\mu, 1/R)|^{2})\frac{(1-\hat{s})\hat{s}+2(1+2\hat{s})\hat{m}{_{\ell}^{2}}}{\hat{s}}\Bigg \},\\
{\cal P}_{yx}&=&\frac{-2}{\Delta}\
Im[C_{9}(\mu, 1/R)C_{10}^\ast(\mu, 1/R)](1-\hat{s})\sqrt{1-\frac{4\hat{m}{_{\ell}^{2}}}{\hat{s}}},\\
{\cal P}_{xy}&=&{\cal P}_{yx},\\
{\cal
P}_{zx}&=&\frac{-3\pi}{2\sqrt{\hat{s}}\Delta}\hat{m}{_{\ell}}\
\Bigg\{2Im[C_{7}(\mu, 1/R)C_{10}^\ast(\mu, 1/R)]+Im[C_{9}(\mu, 1/R)C_{10}^\ast(\mu, 1/R)]\Bigg\},\\
{\cal
P}_{yy}&=&\frac{1}{\Delta}\Bigg\{24Re[C_{7}(\mu, 1/R)C_{9}^\ast(\mu, 1/R)]
\frac{\hat{m}{_{\ell}^{2}}}{\hat{s}}-4(|C_{9}(\mu, 1/R)|^{2}+
|C_{10}(\mu, 1/R)|^{2})\frac{(1-\hat{s})\hat{m}{_{\ell}^{2}}}{\hat{s}}\nonumber
\\&&+(|C_{9}(\mu, 1/R)|^{2}-|C_{10}(\mu, 1/R)|^{2})((-1+\hat{s})+\frac{6\hat{m}{_{\ell}^{2}}}{\hat{s}})\nnb
\\&&+4|C_{7}(\mu, 1/R)|^{2}\frac{((1-\hat{s})\hat{s}
+2(2+\hat{s})\hat{m}{_{\ell}^{2}})}{\hat{s}^{2}}\Bigg\},\\
{\cal P}_{zy}&=&\frac{3\pi}{2\sqrt{\hat{s}}\Delta}\hat{m}{_{\ell}}
\sqrt{1-\frac{4\hat{m}{_{\ell}^{2}}}{\hat{s}}}\\&&
\Bigg\{2Re[C_{7}(\mu, 1/R)C_{10}^\ast(\mu, 1/R)]-|C_{10}(\mu, 1/R)|^{2}+Re[C_{9}(\mu, 1/R)C_{10}^\ast(\mu, 1/R)]\hat{s}\Bigg\},\nnb\\
{\cal P}_{xz}&=&-{\cal P}_{zx},\\
{\cal P}_{yz}&=&\frac{3\pi}{2\sqrt{\hat{s}}\Delta}\hat{m}{_{\ell}}
\sqrt{1-\frac{4\hat{m}{_{\ell}^{2}}}{\hat{s}}}\\&&
\Bigg\{2Re[C_{7}(\mu, 1/R)C_{10}^\ast(\mu, 1/R)]+|C_{10}(\mu, 1/R)|^{2}
+Re[C_{9}(\mu, 1/R)C_{10}^\ast(\mu, 1/R)]\hat{s}\Bigg\},\nnb\\
{\cal
P}_{zz}&=&\frac{1}{2\Delta}\Bigg\{12Re[C_{7}(\mu, 1/R)C_{9}^\ast(\mu, 1/R)]
(1-\frac{2\hat{m}{_{\ell}^{2}}}{\hat{s}})
+\frac{4|C_{7}(\mu, 1/R)|^{2}(2+\hat{s})(1-\frac{2\hat{m}{_{\ell}^{2}}}{\hat{s}})}{\hat{s}}\nonumber
\\&&+(|C_{9}(\mu, 1/R)|^{2}+|C_{10}(\mu, 1/R)|^{2})(1+2\hat{s}-\frac{6(1+\hat{s})\hat{m}{_{\ell}^{2}}}{\hat{s}})\nonumber
\\&&+\frac{2(|C_{9}(\mu, 1/R)|^{2}-|C_{10}(\mu,
1/R)|^{2})(2+\hat{s})\hat{m}{_{\ell}^{2}}}{\hat{s}}\Bigg\}.
\end{eqnarray}

Except ${\cal P}_{zz}$ which is 2 times smaller than the one
obtained in Ref.~\cite{Bensalem:2002ni}, the other ${\cal P}_{ij}$'s
calculated in Ref.~\cite{Bensalem:2002ni} can be achieved by the
replacement of $C_i(\mu, 1/R)\rightarrow C_i^{eff}$ where $i=7, ~9,
~10$. Also, it is obvious that the asymmetries proportional to the
imaginary parts of the Wilson coefficients are small in the SM and
vanish in the ACD where all Wilson coefficients are considered to be
real.

Equipped with the definition of the spin directions of Eq.
(\ref{frame}) in the CM frame of leptons, we can evaluate the
forward-backward asymmetries corresponding to various polarization
components of the $\ell^-$ and/or $\ell^+$ spin by
writing\cite{Bensalem:2002ni}:
\bea
  A_{FB}({\bf s^{+}},{\bf s^{-}},\hat{s})&=& A_{FB}(\hat{s})+ \Big[
{\cal A}^{-}_x s_x^- +{\cal A}^{-}_y s_y^- +{\cal A}^{-}_z s_z^-
+{\cal A}^{+}_x s_x^+ +{\cal A}^{+}_y s_y^+ +{\cal A}^{+}_z s_z^+
\nnb
\\
&& \hskip1truein +~{\cal A}_{xx} s_x^+ s_x^- +{\cal A}_{xy} s_x^+
                s_y^- +{\cal A}_{xz} s_x^+ s_z^- \nnb\\
&& \hskip1truein +~{\cal A}_{yx} s_y^+ s_x^- +{\cal A}_{yy} s_y^+
                s_y^- +{\cal A}_{yz} s_y^+ s_z^- \nnb \\
&& \hskip1truein +~{\cal A}_{zx} s_z^+ s_x^- +{\cal A}_{zy} s_z^+
                s_y^- +{\cal A}_{zz} s_z^+ s_z^- \Big] ~.
\eea
The different polarized forward-backward asymmetries are then
calculated as follows:
\bea
{\cal A}^{+}_x &=& 0, \\
{\cal A}^{+}_y &=& \frac{2}{\Delta}\,{{\rm Re}(C_9(\mu, 1/R)
     C_{10}^\ast(\mu, 1/R))}\,\frac{(1-\hat{s})\,
     \hat{m}_\ell}{\sqrt{\hat{s}}}\,\sqrt{1 -
     \frac{4\,\hat{m}_\ell^2}{\hat{s}}}, \\
{\cal A}^{+}_z &=&\frac{1}{\Delta}\Bigg \{ 6\,{{\rm Re}(C_7(\mu, 1/R)
     C_9^\ast(\mu, 1/R))} - \frac{6\,|C_7(\mu, 1/R)|^2}{\hat{s}}\nnb\\ && -
     3\,(\,|C_9(\mu, 1/R)|^2-|C_{10}(\mu, 1/R)|^2)\,\hat{m}_\ell^2 \nnb\\ && -
     12\,{{\rm Re}(C_7(\mu, 1/R) C_{10}^\ast(\mu, 1/R))}\,\frac{\hat{m}_\ell^2}
     {\hat{s}} - ~6\,{{\rm Re}(C_9(\mu, 1/R)
     C_{10}^\ast(\mu, 1/R))}\,\frac{\hat{m}_\ell^2} {\hat{s}} \nnb \\ &&
     -~\frac{3}{2}\,(\,|C_9(\mu, 1/R)|^2+|C_{10}(\mu, 1/R)|^2)\,\hat{s}\, ( 1 -
     \frac{2\,\hat{m}_\ell^2}{\hat{s}})\Bigg \}, \\
{\cal A}^{-}_x &=& 0,\\
{\cal A}^{-}_y &=& {\cal A}^{+}_y, \\
{\cal A}^{-}_z &=&\frac{1}{\Delta}\Bigg \{ -6\,{{\rm Re}(C_7(\mu, 1/R)
     C_9^\ast(\mu, 1/R))} - \frac{6\,|C_7(\mu, 1/R)|^2}{\hat{s}} \nnb\\ &&-
     3\,(\,|C_9(\mu, 1/R)|^2-|C_{10}(\mu, 1/R)|^2)\,\hat{m}_\ell^2 \nnb\\ &&
     +~12\,{{\rm Re}(C_7(\mu, 1/R) C_{10}^\ast(\mu, 1/R))}\,\frac{\hat{m}_\ell^2}
     {\hat{s}} +~6\,{{\rm Re}(C_9(\mu, 1/R)
     C_{10}^\ast(\mu, 1/R))}\,\frac{\hat{m}_\ell^2} {\hat{s}} \nnb \\ &&
     -~\frac{3}{2}\,(\,|C_9(\mu, 1/R)|^2+|C_{10}(\mu, 1/R)|^2)\,\hat{s}\, ( 1 -
     \frac{2\,\hat{m}_\ell^2}{\hat{s}})\Bigg \}, \\
{\cal A}_{xx} &=& 0,\\
{\cal A}_{xy} &=& \frac{-6}{\Delta}\,( 2\,{{\rm
Im}(C_7(\mu, 1/R)\,C_{10}^\ast(\mu, 1/R))} + {{\rm
Im}(C_9(\mu, 1/R)\,C_{10}^\ast(\mu, 1/R))})\,\frac{{{\hat{m}_\ell}}^2}{\hat{s}}, \\
{\cal A}_{xz} &=& \frac{2}{\Delta}\,{{\rm
     Im}(C_9(\mu, 1/R)C_{10}^\ast(\mu, 1/R))}\,\frac{(1-\hat{s})\,{\hat{m}_\ell}
     }{\sqrt{\hat{s}}}\,\sqrt{1 -
     \frac{4\,\hat{m}_\ell^2}{\hat{s}}},
     \\
{\cal A}_{yx} &=& -{\cal A}_{xy},\\
{\cal A}_{yy} &=& 0,\\
{\cal A}_{yz}\! &=& \!\Big( 2
      |C_9(\mu, 1/R)|^2-\frac{8\,|C_7(\mu, 1/R)|^2}{\hat{s}} \Big) \, \frac{(
      1 - \hat{s})\,\hat{m}_\ell}{\Delta\sqrt{\hat{s}}}, \\
{\cal A}_{zx} &=& {\cal A}_{xz},\\
{\cal A}_{zy} &=& {\cal A}_{yz}, \\
{\cal A}_{zz} &=& \frac{-3}{\Delta}\,( 2\,{{\rm
     Re}(C_7(\mu, 1/R)\,C_{10}^\ast(\mu, 1/R))} \nnb\\ &&+ {{\rm
     Re}(C_9(\mu, 1/R)\,C_{10}^\ast(\mu, 1/R))}\,\hat{s}) \, {\sqrt{1 -
     \frac{4\,{{\hat{m}_\ell}}^2}{\hat{s}}}} ~.
\eea

Here, ${\cal A}_{zz}$ coincides with $-{\cal A}_{FB}$ in the SM and
any of its extensions\cite{Bensalem:2002ni}, provided that the
operator structure remains the same. In other words,  a significant
difference between ${\cal A}_{zz}$ and ${\cal A}_{FB}$ happens when
the new type of interactions are taken into account in the effective
Hamiltonian, i.e., the tensor type and scalar type interactions
differ between  ${\cal A}_{zz}$ and ${\cal
A}_{FB}$\cite{Aliev:2004hi}.

Note that,  ${\cal A}_{ij}$ coefficients calculated in
Ref.~\cite{Bensalem:2002ni} can again be obtained by the replacement
of $C_i(\mu, 1/R)\rightarrow C_i^{eff}$ where $i=7, ~9, ~10$.
 \section{Numerical analysis}
In this section, we study the dependence of the double-lepton
polarization and the polarized lepton FB asymmetries on the
compactification parameters($1/R)$. We use the SM parameters shown
in Table 1:
\begin{table}[h]
        \begin{center}
        \begin{tabular}{|l|l|}
        \hline
        \multicolumn{1}{|c|}{Parameter} & \multicolumn{1}{|c|}{Value}     \\
        \hline \hline
         $\alpha_{s}(m_Z)$                   & $0.119$  \\
        $\alpha_{em}$                   & $1/129$\\
        $m_{W}$                   & $80.41$ (GeV) \\
        $m_{Z}$                   & $91.18$ (GeV) \\
        $sin^2(\theta_W)     $      & $0.223$  \\
        $m_{b}$                   & $4.7$ (GeV) \\
        $m_{\mu}$                   & $0.106$ (GeV) \\
        $m_{\tau}$                  & $1.780$ (GeV) \\
        \hline
        \end{tabular}
        \end{center}
\caption{The values of the input parameters used in the numerical
          calculations.}
\label{input}
\end{table}

The allowed range in the ACD model for the Wolfenstein parameters
shows a small discrepancy in terms of $1/R$  with respect to the SM
values\cite{Buras1}.

The physical observables depend on compactification radius (R) and
$\hat s$. The conservation of KK parity $(-1)^j$, with $j$ as the KK
number, implies the absence of tree-level contribution of KK states
at the low energy regime. This allows us to establish a bound:
$1/R>250~GeV$ by the analysis of Tevatron run I data\cite{ACD}. The
same bound can be obtained by the analysis of measured branching
ratio of $B\to X_s \gamma$ decay\cite{Buras1, Buras2}. A sharper
constraint on $1/R$  is established by taking into account the
leading order contributions due to the exchange of Kaluza-Klein
modes as well as the available next-to-next-to-leading order
corrections to the branching ratio of $B\to X_s \gamma$ decay
\cite{Haisch}. In what follows, we consider $200<1/R<1000~GeV$.
Furthermore, in order to do two-dimensional analysis about the
observables, we must eliminate one of the variables either $1/R$ or
$\hat s$. We do two types of analysis, first, we choose fixed values
of the $1/R\sim\{200, 350, 500\}~GeV$ and look at the $\hat s $
dependency of the FB asymmetries. Not that, zero point position of
the FB asymmetries in terms of the $\hat s$ is less sensitive to the
hadronic uncertainties in exclusive decay channels. Second, we
eliminate the $\hat s$ dependency from double-lepton polarization
asymmetries by performing integration over $\hat{s}$ in the allowed
region, i.e., we consider the averaged values of the various
asymmetries. The average gained over $\hat{s}$ is defined as: \bea
\la {\cal{P}} \ra = \frac{\ds \int_{4
\hat{m}_\ell^2}^{(1-\sqrt{\hat{r}_K})^2} {\cal{P}} \frac{d{\cal
B}}{d \hat{s}} d \hat{s}} {\ds \int_{4
\hat{m}_\ell^2}^{(1-\sqrt{\hat{r}_K})^2} \frac{d{\cal B}}{d \hat{s}}
d \hat{s}}~.\nnb \eea Our quantitative analysis indicates that some
of the observables are less sensitive to the $1/R$; i.e., the
maximum deviations from the SM are $\sim 1\%$. We do not present
those dependencies on the $1/R$ with relevant figures. We present
our analysis for strongly dependent functions in a series of
figures. We do not present some of the observables where the SM and
ACD values are almost vanishing or their deviation with respect to
the SM values are negligible (less than $1\%$).

From these figures, we deduce the following results:

\subsection{ Differential polarized FB asymmetries}
Figures 1--7 depict the $\hat s$ dependency of the single or
double-lepton polarization FB asymmetries for three fixed value of
the $1/R=200;350;500~GeV$.
\begin{itemize}
\item{${\cal A}_y^+(\hat s)$ for $\mu$ channel  depicts strong dependency in the nonresonance region
where $\hat s\sim\{0.0-0.02\}$. The magnitude of ${\cal A}_y^+$ is
enhanced by decreasing the compactification scale $1/R$ for $\mu$
channel (see Fig. 1). For high $\hat s$ region, which is also a
nonresonance region, the discrepancy almost vanishes. }

\item{ ${\cal A}_z^-(\hat s)$ for $\tau$ lepton and
${\cal A}_z^+(\hat s)$ for $\mu$ lepton are suppressed in the ACD
model. In particular, when compactification scale $1/R$ is
decreased, the deviation corresponding to the SM values is also
decreased (see Figs. 2, 3).}

\item{${\cal A}_z^+(\hat s)$ for the $\tau$
channel depicts almost homogenous discrepancy with respect to the
 SM values in all kinematically allowed regions.
While the SM values are always negative, considering the ACD model,
it can get positive values at lower momentum transfer region (see
Fig. 4).
 A measurement of sign of this observable can be used in general to
 probe
the NP effects, in particular, to test the ACD model. }

\item{ The zero point position and the magnitude of ${\cal A}_{yz}(\hat s)$ for the $\mu$ channel
 is almost in sensitive to the new parameter of the ACD model, i.e.,
the compactification scale $1/R$ (see Fig. 5).}

\item{ ${\cal A}_{yz}(\hat s)$ for the $\tau$ channel shows strong dependency
 at lower momentum transfer region.
 The magnitude of ${\cal A}_{yz}(\hat s)$ is enhanced
by decreasing the compactification scale $1/R$ (see Fig. 6). For
high $\hat s$ region, which is also a nonresonance region, the
discrepancy almost vanishes.}

\item{The zero point position of ${\cal A}_{zz}(\hat s)$ for the $\mu$
channel is shifted to left of the SM point. Also, we find that
${\cal A}_{zz}(\hat s)$ coincides with unpolarized FB asymmetries
${\cal A}_{FB}(\hat s)$[see Fig. 7a], which is  given in
Ref.~\cite{Buras2}. The zero point position of ${\cal A}_{zz}(\hat
s)~(s_0=2C_7/C_9)$ is sensitive to the $C_7/C_9$. We show that $s_0$
is increasing function of $1/R$. The sizable deviation from the
corresponding SM value of $s_0$ occurs at the small values of
$1/R$[see Fig.7b]. This point is especially important for the
exclusive decays where the hadronic uncertainty almost vanishes at
this point. }
\end{itemize}

\subsection{ Averaged Double-Lepton Polarization Asymmetries}
\begin{itemize}
\item{ Taking into account the ACD model,  the magnitudes  of all
double-lepton polarization asymmetries ($\la{\cal P}_{ij} \ra$),
except $\la{\cal P}_{zz} \ra$, are enhanced. Moreover, the
discrepancy is sizable for smaller values of compactification scale
$1/R$ (see Figs. 8--14). On the other hand, the $\la{\cal P}_{zz}
\ra$ is suppressed in the ACD model. Measurements of magnitude and
sign of these observables are good tools to search for physics
beyond the SM, in particular, to look for the UED.}
\end{itemize}
Finally, some remarks are in order:

 First, the quantitative
estimations about the accessibility to measure
 the various physical observables are important issues from the
 experimental point of view.
A required number of  $B \bar{B}$ pairs in terms of the branching
ratio ${\cal B}$ at $n \sigma$ level, the efficiencies of the
leptons $s_1$ and $s_2$, and various asymmetry functions are given
as: \bea N = \frac{n^2}{{\cal B} s_1 s_2 \la {\cal A} \ra^2}~,\nnb
\eea where ${\cal A}$ can be an asymmetry.

The efficiencies of detection of the $\tau$--leptons range from
$50\%$ to $90\%$ for their various decay modes\cite{R6016}. Also,
the error in $\tau$--lepton polarization is estimated to be about
$10\%- 15\%$ \cite{R6017}. So, the error in measurement of the
$\tau$--lepton asymmetries is approximately $20\%-30\%$, and the
error in obtaining the number of events is about $50\%$. It can be
understood that in order to detect the asymmetries in the $\mu$ and
$\tau$ channels at the $3\sigma$ level with the asymmetry of ${\cal
A}=1\%$ and efficiency of $\tau \sim 0.5$), the minimum number of
required events are $N\sim 10^{10}$ and $N\sim 10^{11}$ for $\mu$
and $\tau$ leptons, respectively.

On the other hand, the number of $B \bar{B}$ pairs produced at  LHC
are expected to be about $\sim 10^{12}$. Therefore, a typical
asymmetry of (${\cal A}=1\%$) is detectable at LHC. More about these
experimental observables can be found in\cite{Bensalem:2002ni}.

Second, we should note that one can reexamine the constraint  by
studying the branching ratio of pureleptonic B decays and the zero
point position of forward--backward asymmetry of $B\to K^\ast
\ell^+\ell^-$ decay. The former is relatively clean and the latter
point  is almost free of hadronic uncertainties($\sim5\%$). After
having experimental data on pure leptonic B decay and zero point
position of FB, we will obtain the better result for the combination
of $C_9$ and $C_{10}$ from pureleptonic B decay and for the
$C_7/C_9$ from zero point position of FB. In order to fix the Wilson
coefficients and the structure of the interaction Hamiltonian, we
need to study many other observables. For instance, we introduce
some of those observables in the present study, which are polarized
FB and lepton polarization asymmetries.

Finally, the important issue is to find the evidence and to
distinguish the UED model from the others. Even though there are few
studies trying to shed light and introduce a way to distinguish the
UED from the other models (see for example~\cite{Anirban}), this
issue is the common problem of all models and nobody has yet found a
clear solution.
\section{Conclusion}
To sum up, we presented the various asymmetries in inclusive $b
\rar s \ell^+ \ell^-$ transition in the ACD model. The result we obtained are:
\begin{itemize}

\item{The zero point position of double--lepton polarization FB asymmetry in
the ACD model is shifted to the left of the SM position.}

 \item{ Some of the
double--lepton polarization and polarized double or single-lepton
polarization forward--backward which are already accessible at LHC,
depict the strong dependency $(\hat s)$ on the free parameter of the
ACD model, which is compactification scale $R$.}

Thus, the measurement of zero point position of polarized FB
asymmetry as well as  sign or magnitude  polarized FB and
double-lepton polarization asymmetries can serve as a good test
for the predictions of the ACD.
\end{itemize}

\section{Acknowledgment}
The authors would like to thank T. M. Aliev and A. Ozpineci for
their useful discussions. V. Bashiry thanks the Ministry of
Education of Turkish Republic of Northern Cyprus for their partially
support.  Also, M. B. and K. A. would like to thank TUBITAK, Turkish
Scientific and Research Council, for their financial support
provided under the Project No.103T666.

\newpage

\section*{Figure captions}

{\bf Fig. (1)} The dependence of the ${\cal A}_y^+(\hat s)$ of $b\rar
s \mu^+\mu^-$ on  $\hat{s}$  for $1/R=200;350;500~GeV$.\\ \\
{\bf Fig. (2)} The dependence of the ${\cal A}_z^-(\hat s)$ of
$b\rar
s \tau^+\tau^-$ on $\hat{s}$  for $1/R=200;350;500~GeV$.\\ \\
{\bf Fig. (3)} The same as in Fig. (1), but for the ${\cal A}_z^+(\hat s)$.\\ \\
{\bf Fig. (4)} The same as in Fig. (3), but for the $\tau$ lepton.\\ \\
{\bf Fig. (5)} The dependence of the ${\cal A}_{yz}(\hat s)$ of $b\rar
s \mu^+\mu^-$ on  $\hat{s}$  for $1/R=200;350;500~GeV$.\\ \\
{\bf Fig. (6)} The same as in Fig. (5), but for the $\tau$ lepton.\\ \\
{\bf Fig. (7a)} The dependence of the ${\cal A}_{zz}(\hat s)$ of
$b\rar
s \mu^+\mu^-$ on  $\hat{s}$  for $1/R=200;350;500~GeV$.\\ \\
{\bf Fig. (7b)} The dependence of the $s_0$ of $b\rar
s \mu^+\mu^-$ on   $1/R$.\\ \\
{\bf Fig. (8)} The dependence of the $\la{\cal P}_{xx}\ra$ of
$b\rar
s \mu^+\mu^-$ on   $1/R$.\\ \\
{\bf Fig. (9)} The same as in Fig. (8), but for the $\tau$ lepton.\\ \\
{\bf Fig. (10)} The dependence of the $\la{\cal P}_{yy}\ra$ of $b\rar
s \mu^+\mu^-$ on   $1/R$.\\ \\
{\bf Fig. (11)} The same as in Fig. (10), but for the $\tau$ lepton.\\ \\
{\bf Fig. (12)} The dependence of the $\la{\cal P}_{yz}\ra$ of $b\rar
s \tau^+\tau^-$ on   $1/R$.\\ \\
{\bf Fig. (13)} The dependence of the $\la{\cal P}_{zy}\ra$ of $b\rar
s \tau^+\tau^-$ on   $1/R$.\\ \\
{\bf Fig. (14)} The dependence of the $\la{\cal P}_{zz}\ra$ of $b\rar
s \tau^+\tau^-$ on   $1/R$.\\ \\
\newpage
\begin{figure}
\vskip 1.5 cm
    \includegraphics{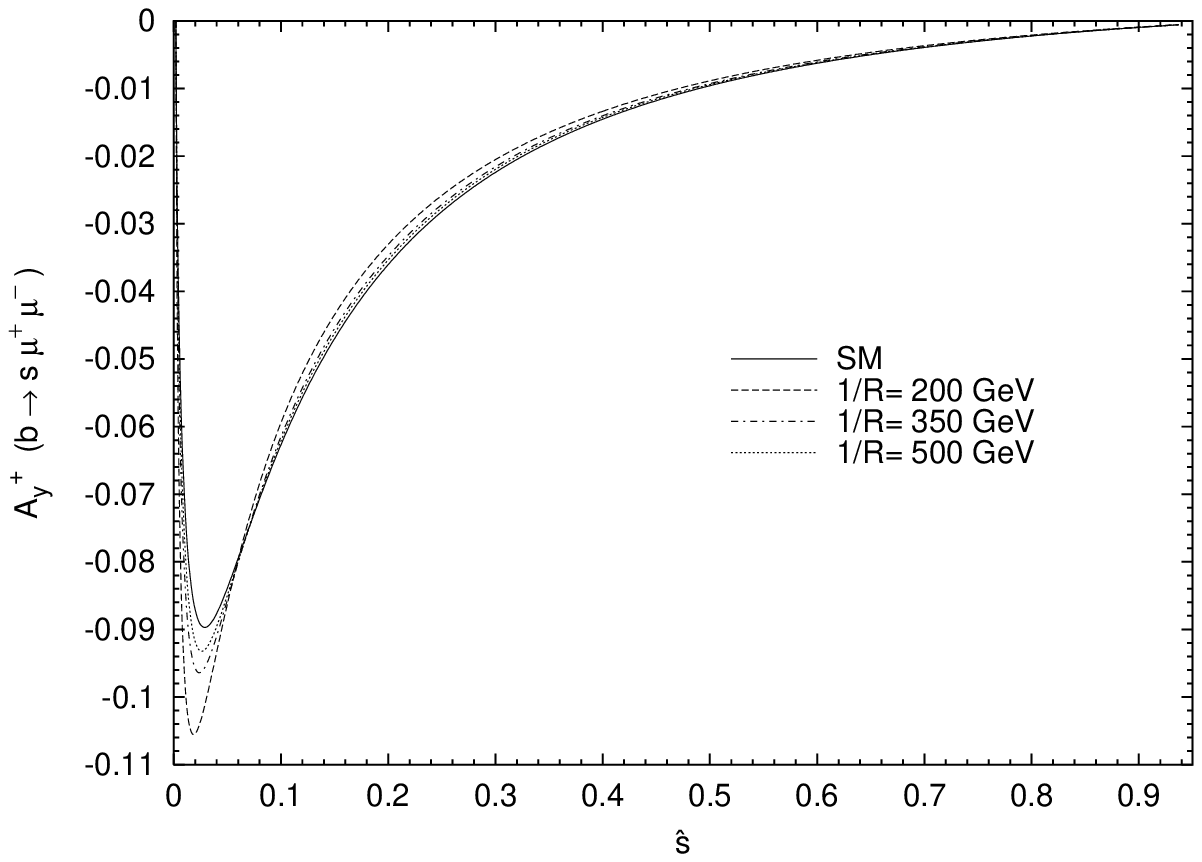}
\vskip 7.8cm \caption{}
\end{figure}

\begin{figure}
\vskip 1.5 cm
    \includegraphics{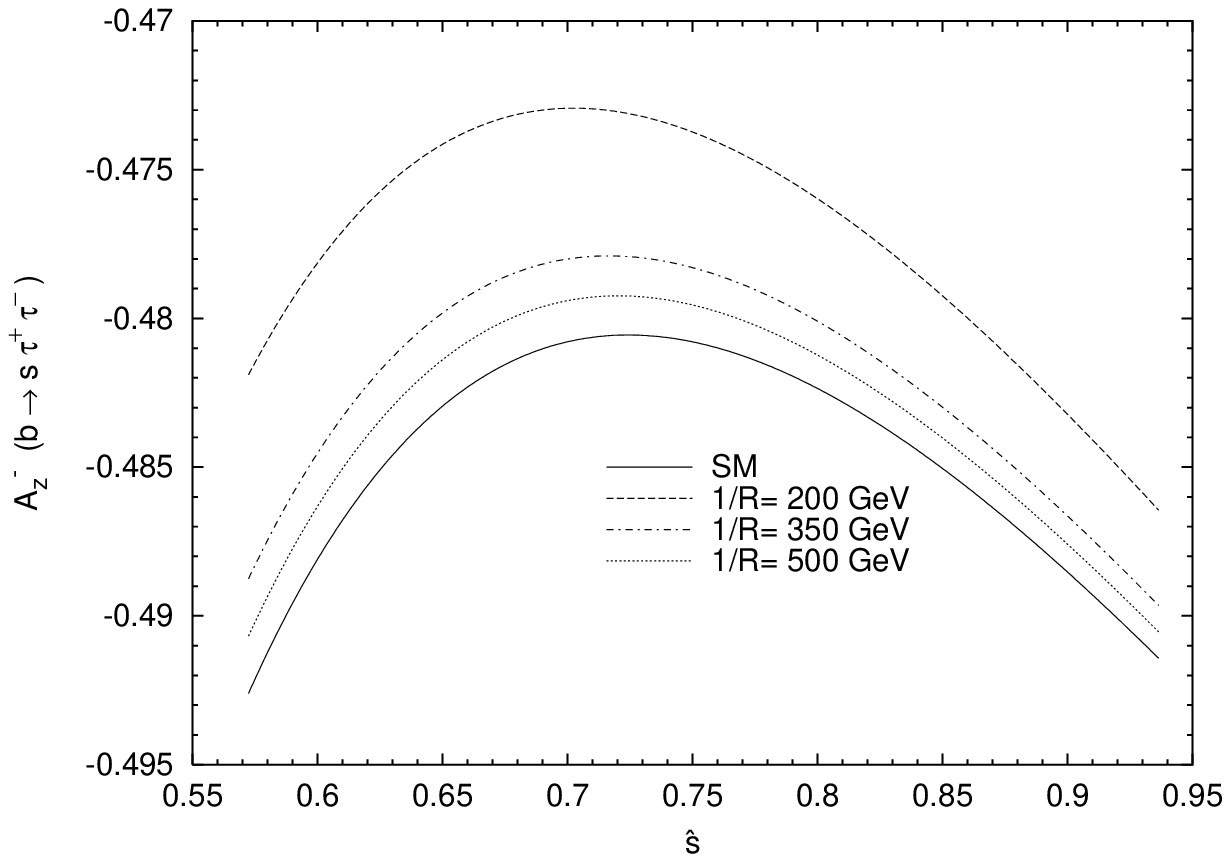}
\vskip 5.8 cm \caption{}
\end{figure}

\begin{figure}
\vskip 2.5 cm
    \includegraphics{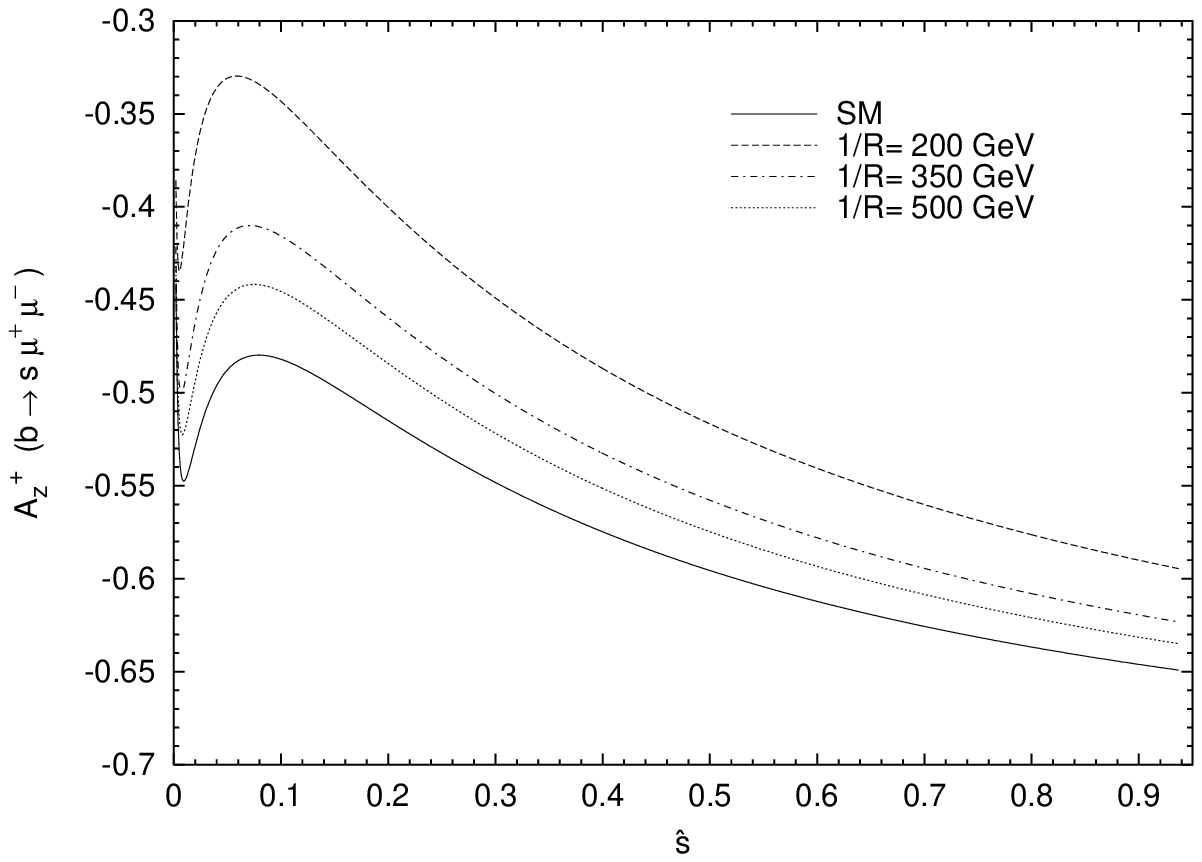}
\vskip 7.8cm \caption{}
\end{figure}

\begin{figure}
\vskip 1.5 cm
    \includegraphics{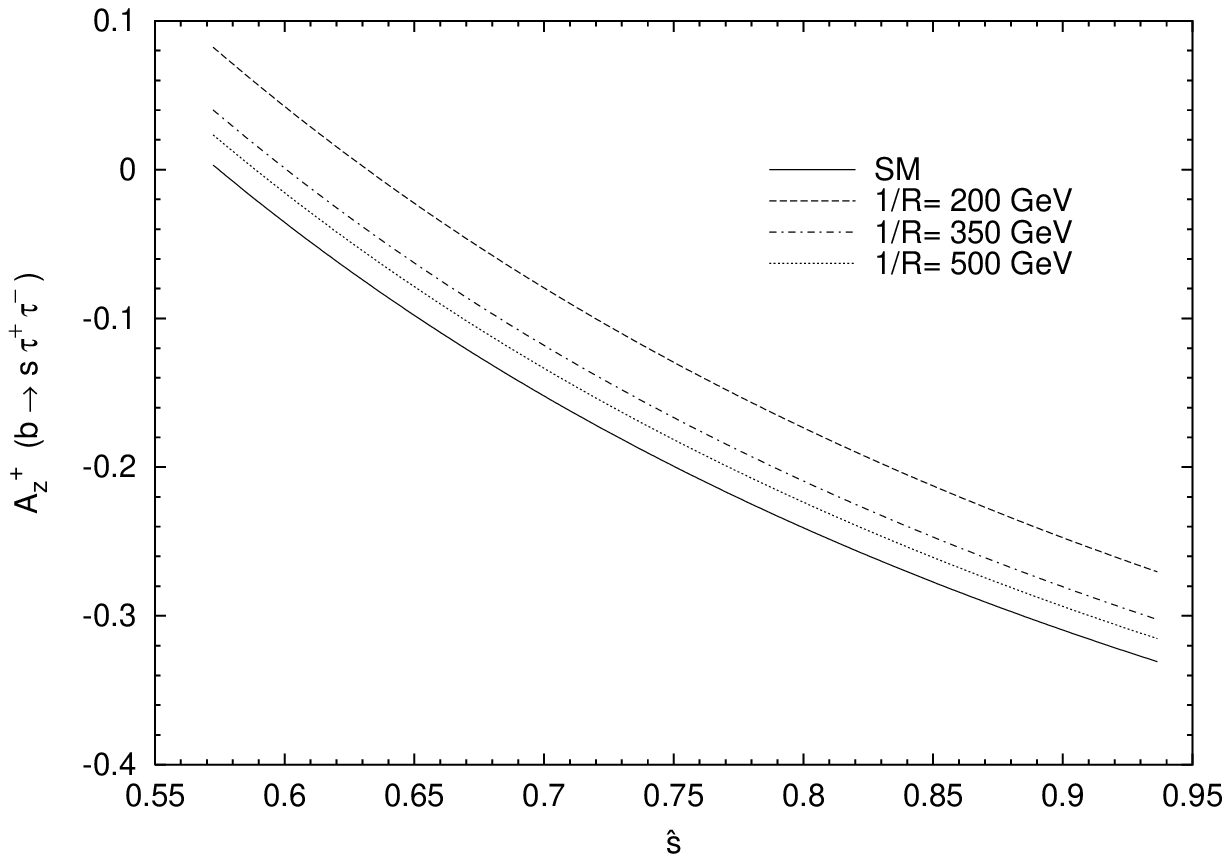}
\vskip 5.8 cm \caption{}
\end{figure}

\begin{figure}
\vskip 2.5 cm
    \includegraphics{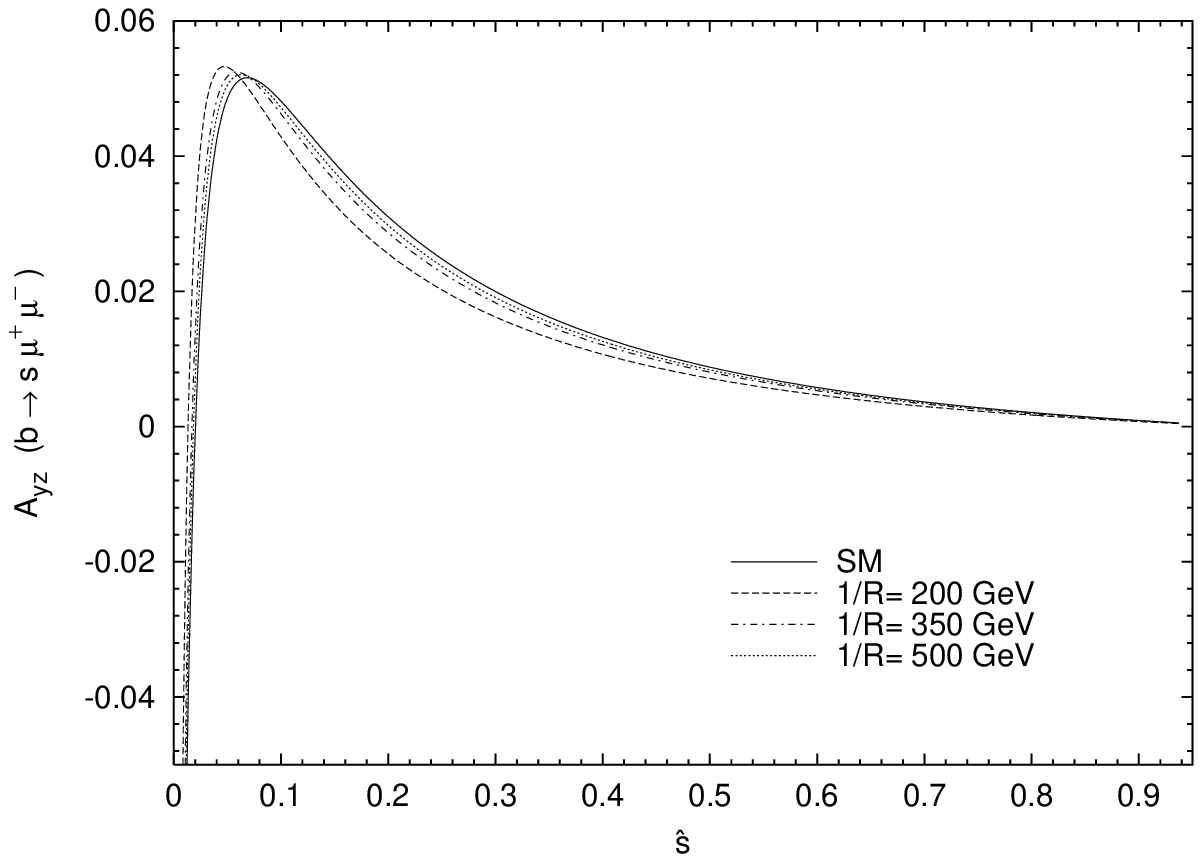}
\vskip 7.8cm \caption{}
\end{figure}

\begin{figure}
\vskip 1.5 cm
    \includegraphics{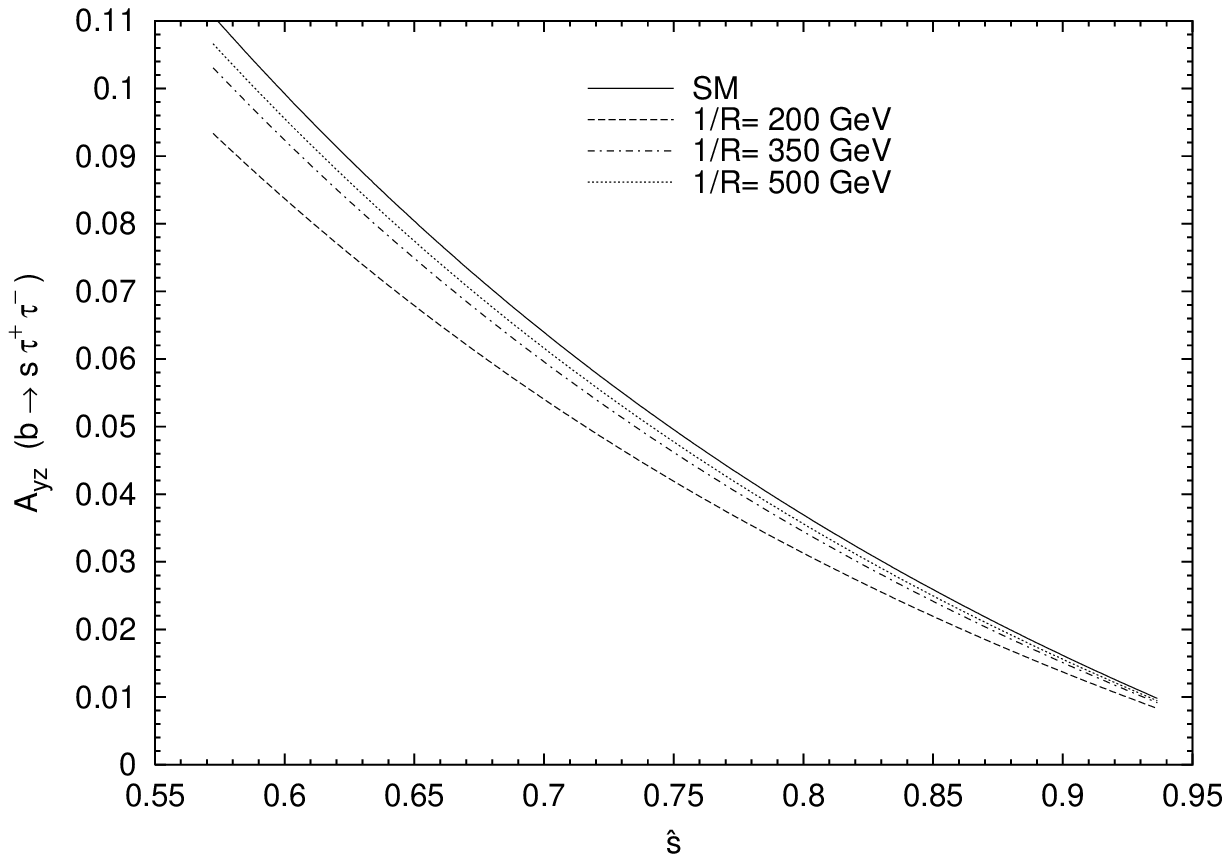}
\vskip 5.8 cm \caption{}
\end{figure}
\begin{figure}
\vskip 0.5 cm
    \includegraphics{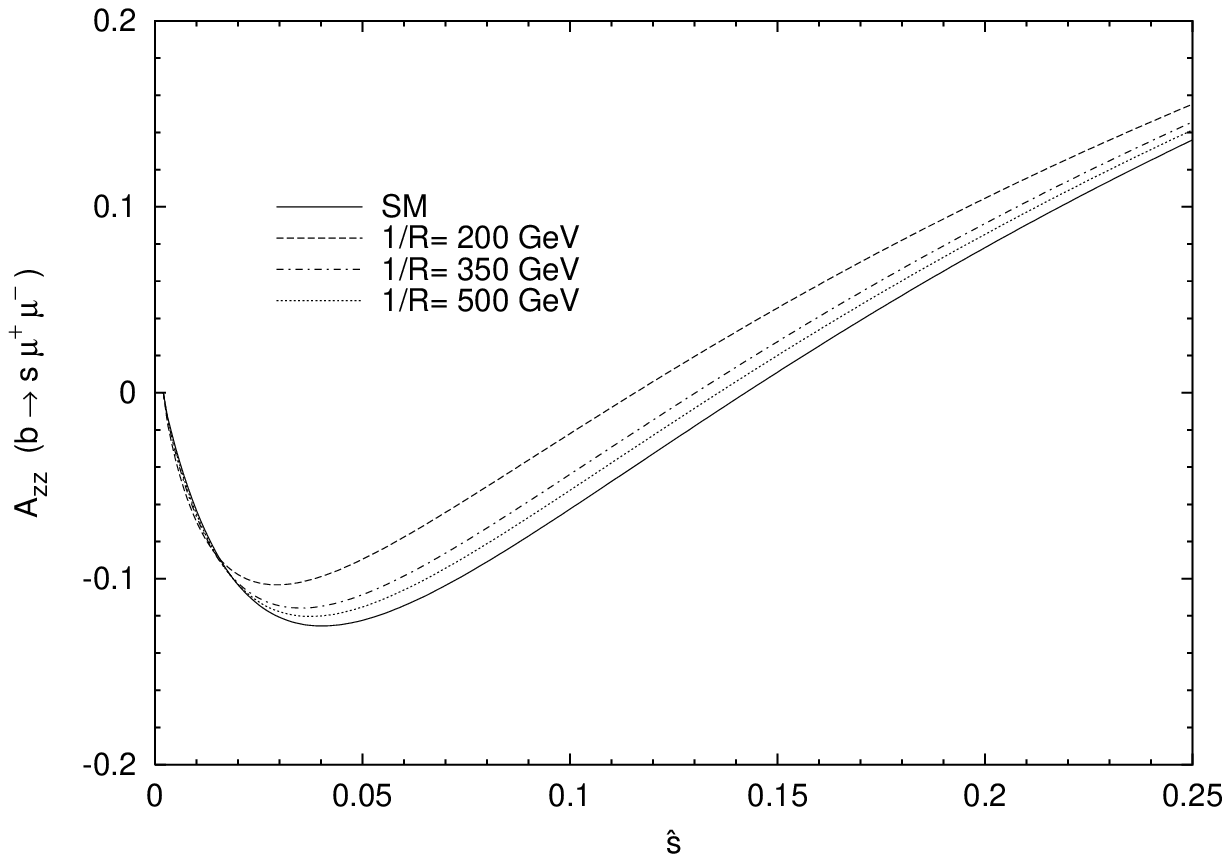}
\includegraphics{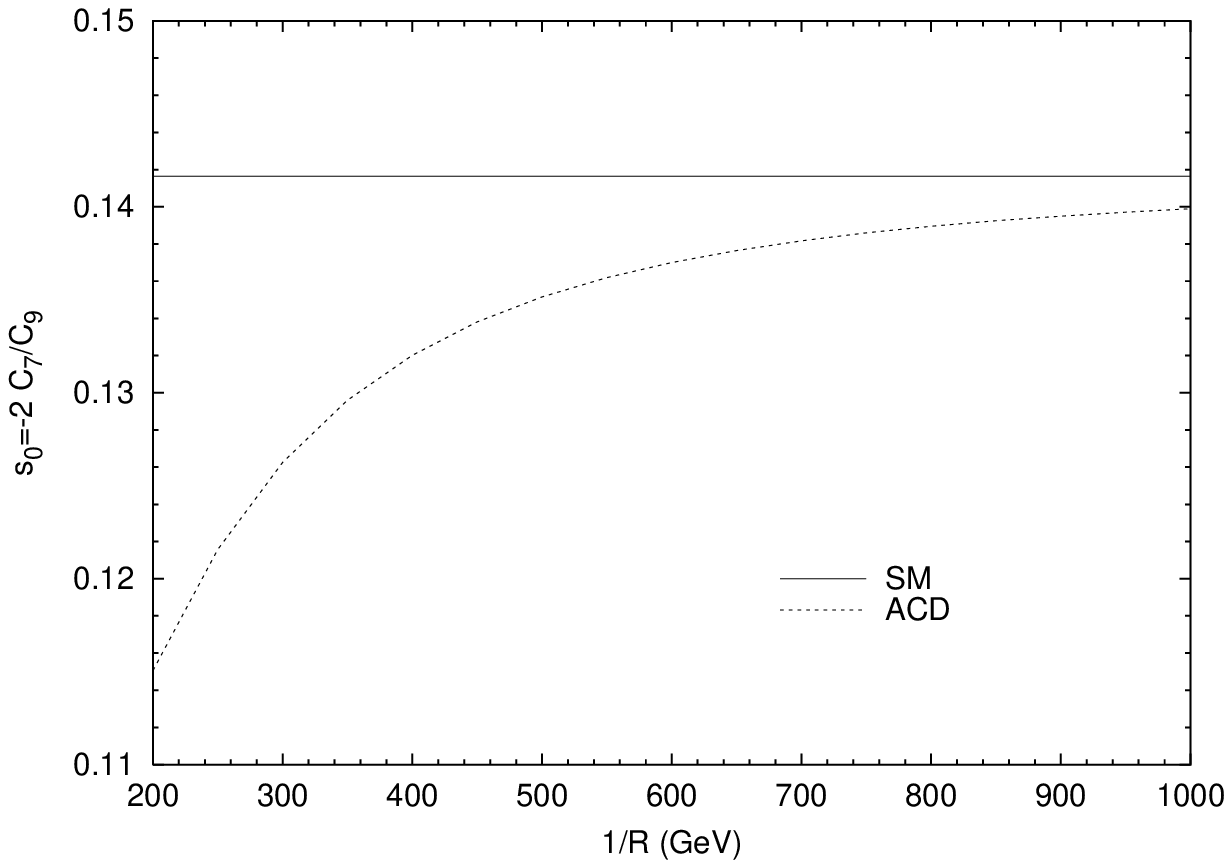}   \vskip 8.8
cm{\small~~~~~~~~~~~~~~~~~~~~~~~~~(
a)~~~~~~~~~~~~~~~~~~~~~~~~~~~~~~~~~~~~~~~~~~~~~~~~~~~~~~~~(
b)}\caption{}
 \end{figure}

\begin{figure}
\vskip 1.5 cm
    \includegraphics{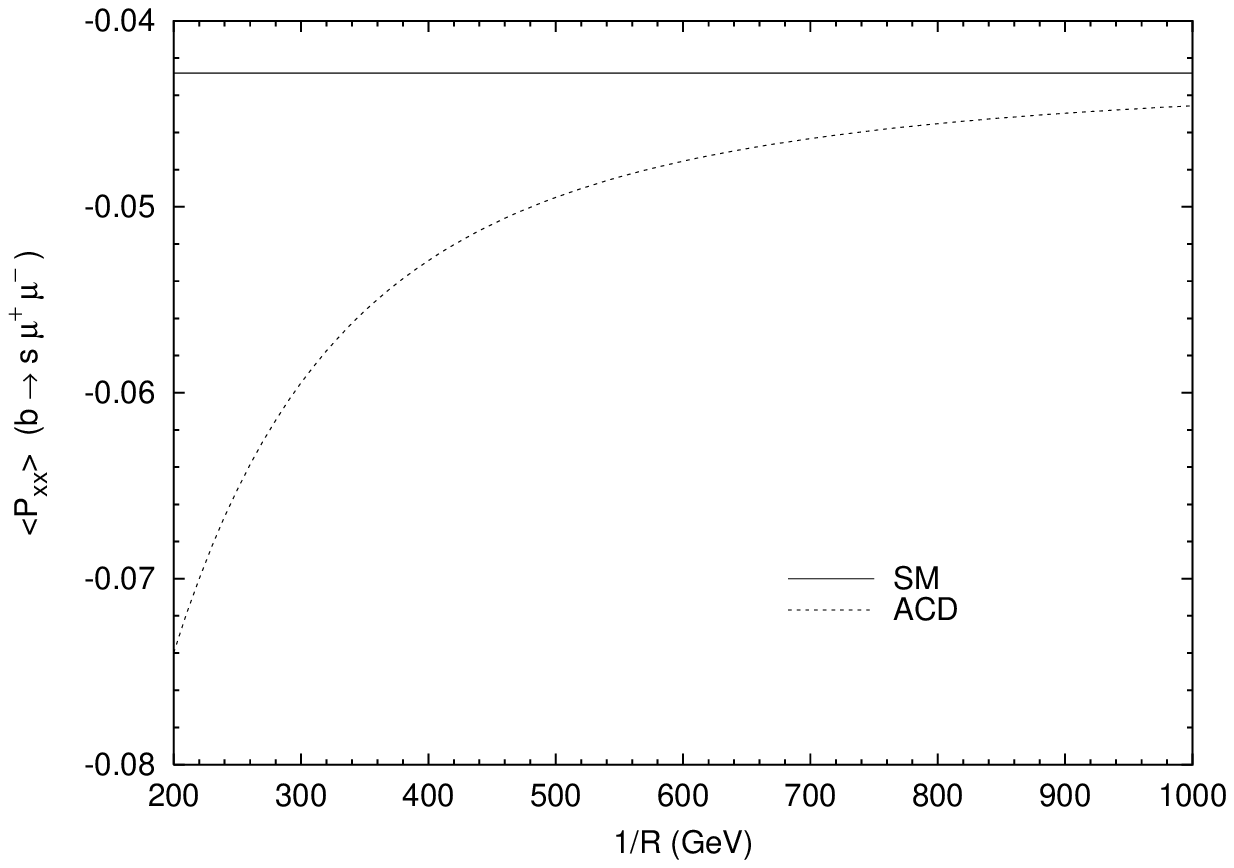}
\vskip 7.8 cm \caption{}
\end{figure}

\begin{figure}
\vskip 2.5 cm
    \includegraphics{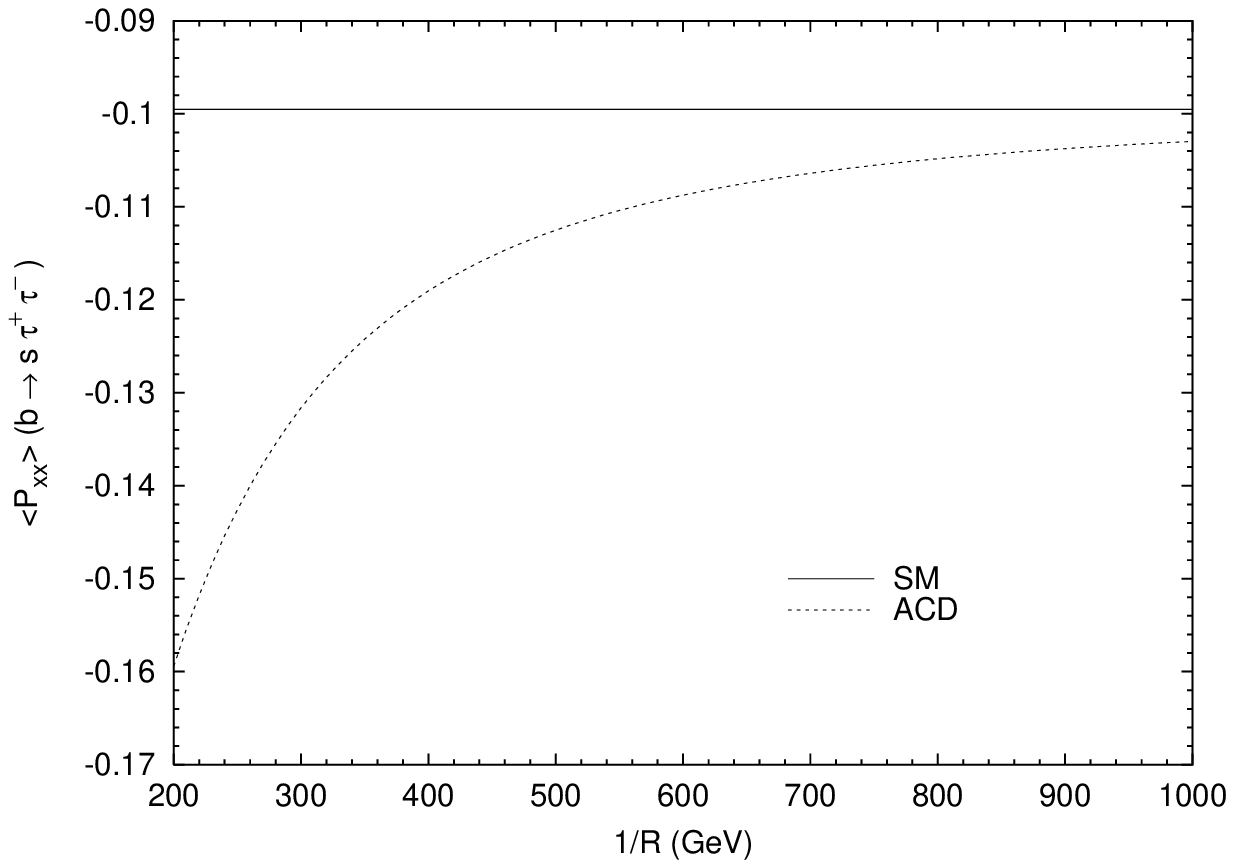}
\vskip 5.8cm \caption{}
\end{figure}

\begin{figure}
\vskip 1.5 cm
    \includegraphics{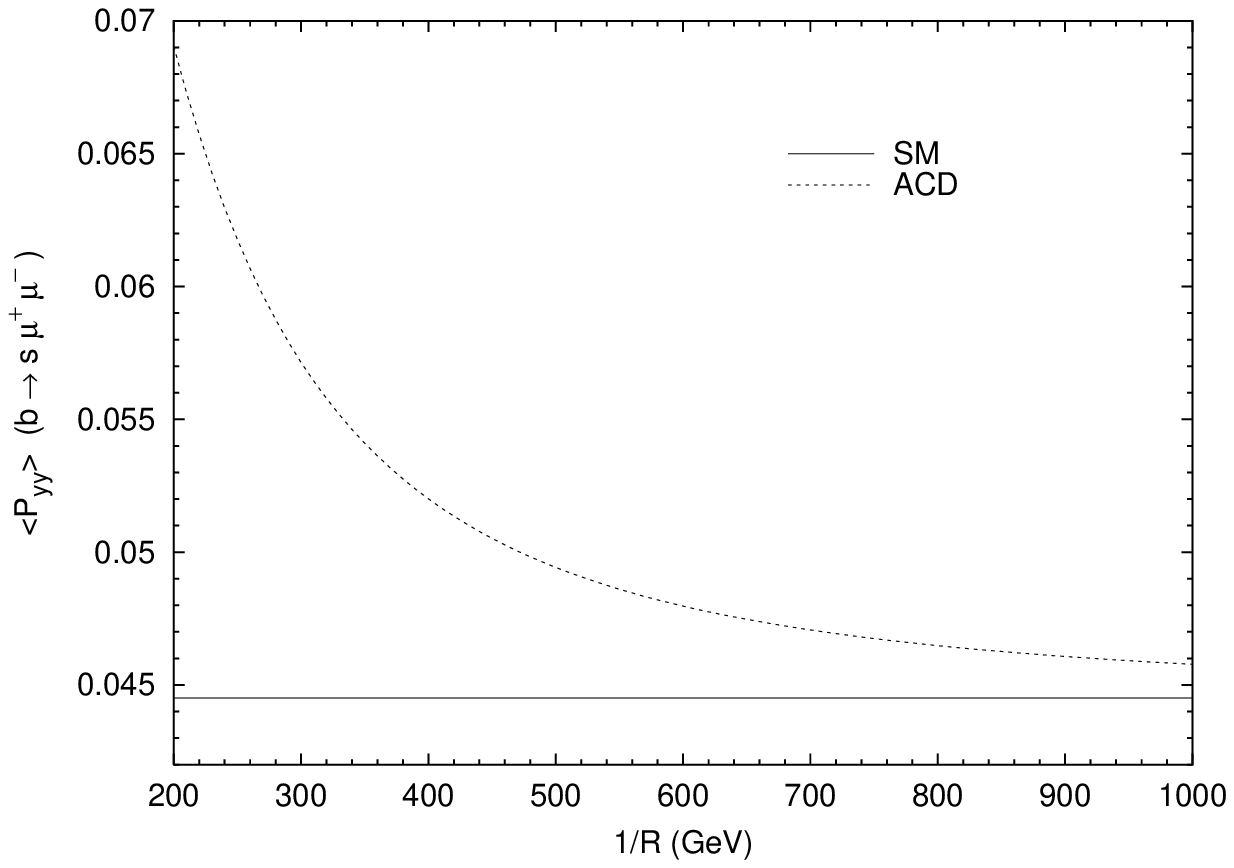}
\vskip 7.8 cm \caption{}
\end{figure}

\begin{figure}
\vskip 1.5 cm
    \includegraphics{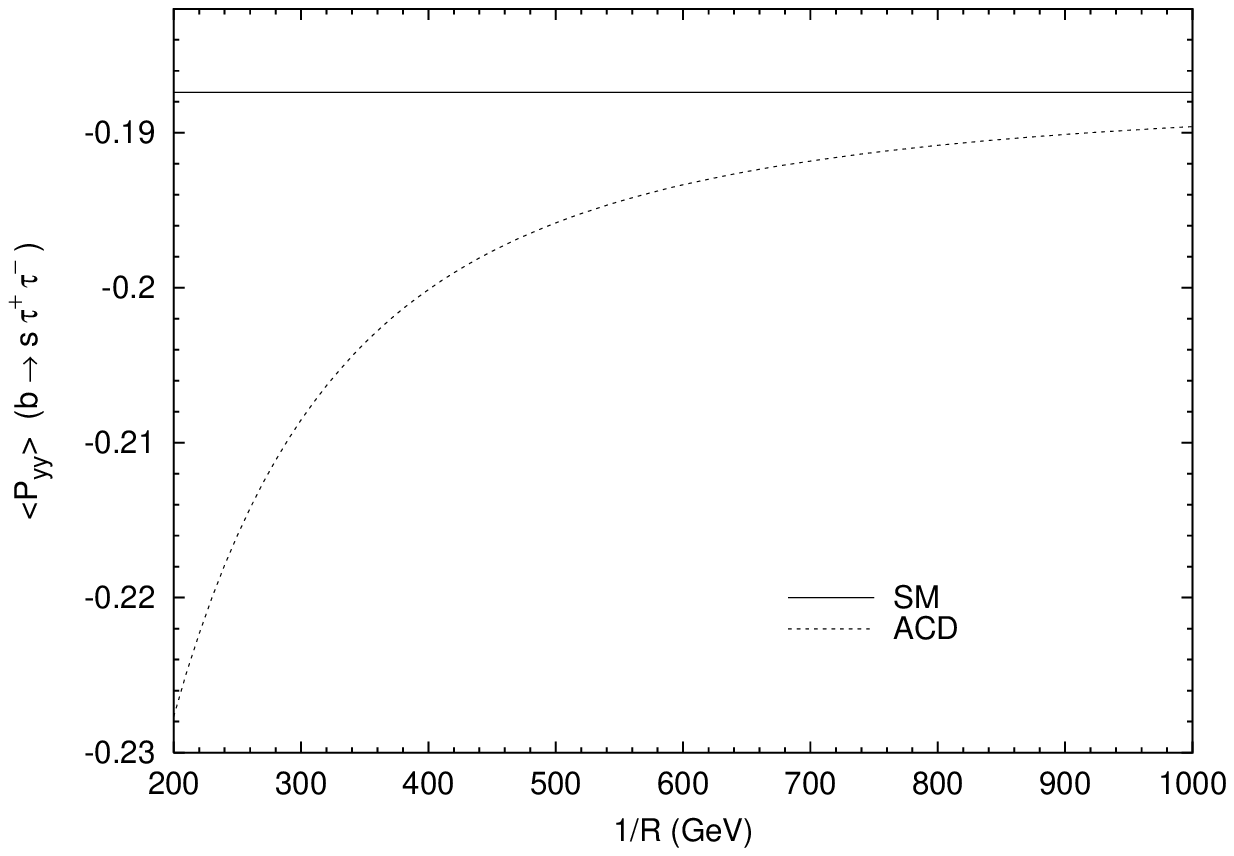}
\vskip 5.8cm \caption{}
\end{figure}

\begin{figure}
\vskip 1.5 cm
    \includegraphics{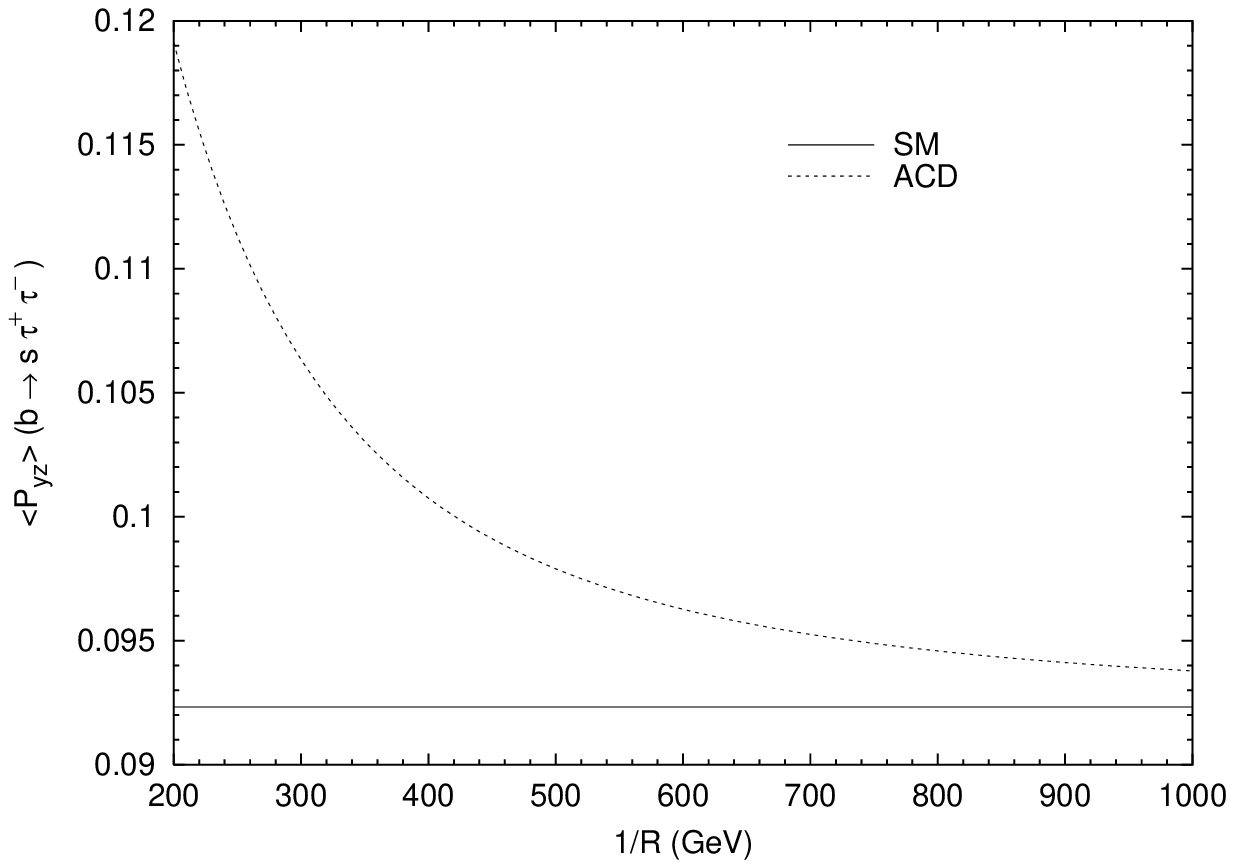}
\vskip 7.8 cm \caption{}
\end{figure}

\begin{figure}
\vskip 1.5 cm
    \includegraphics{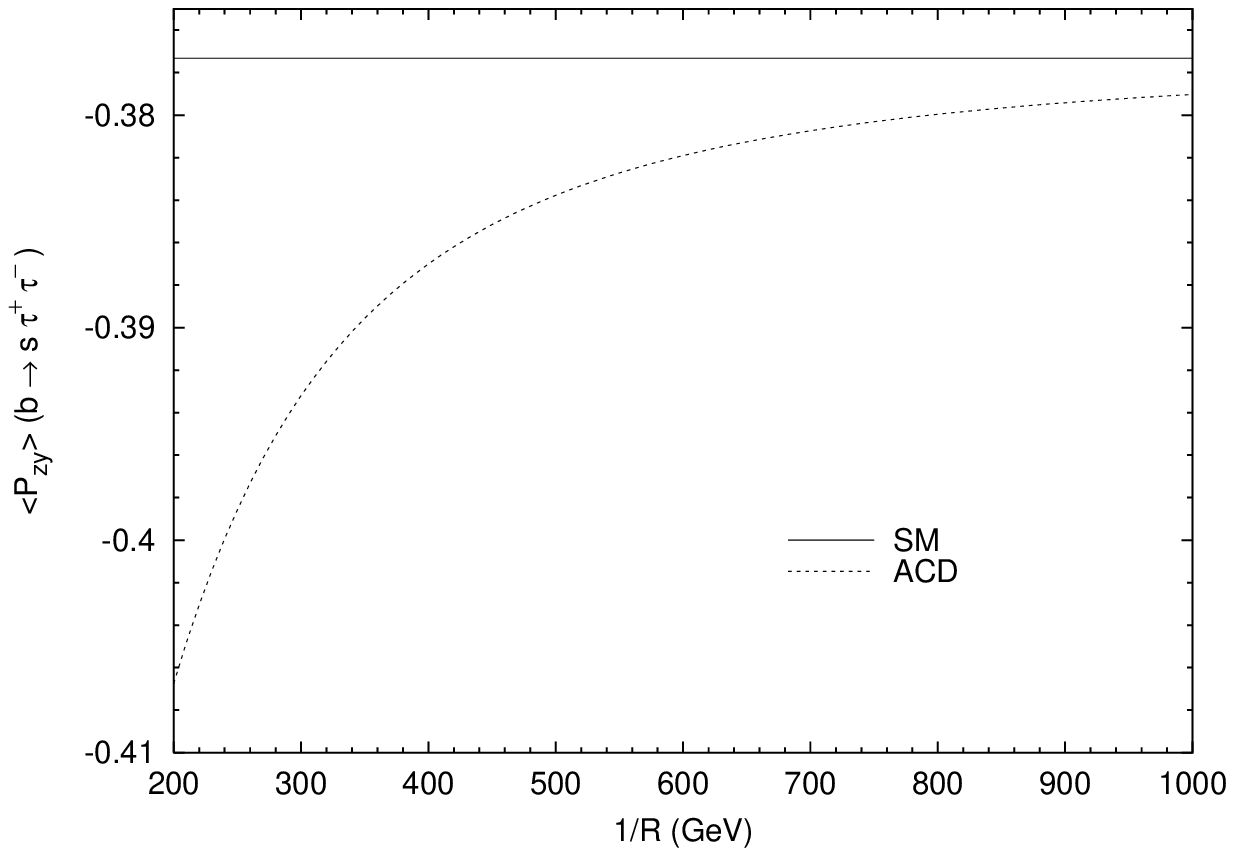}
\vskip 5.8cm \caption{}
\end{figure}

\begin{figure}
\vskip 2.5 cm
    \includegraphics{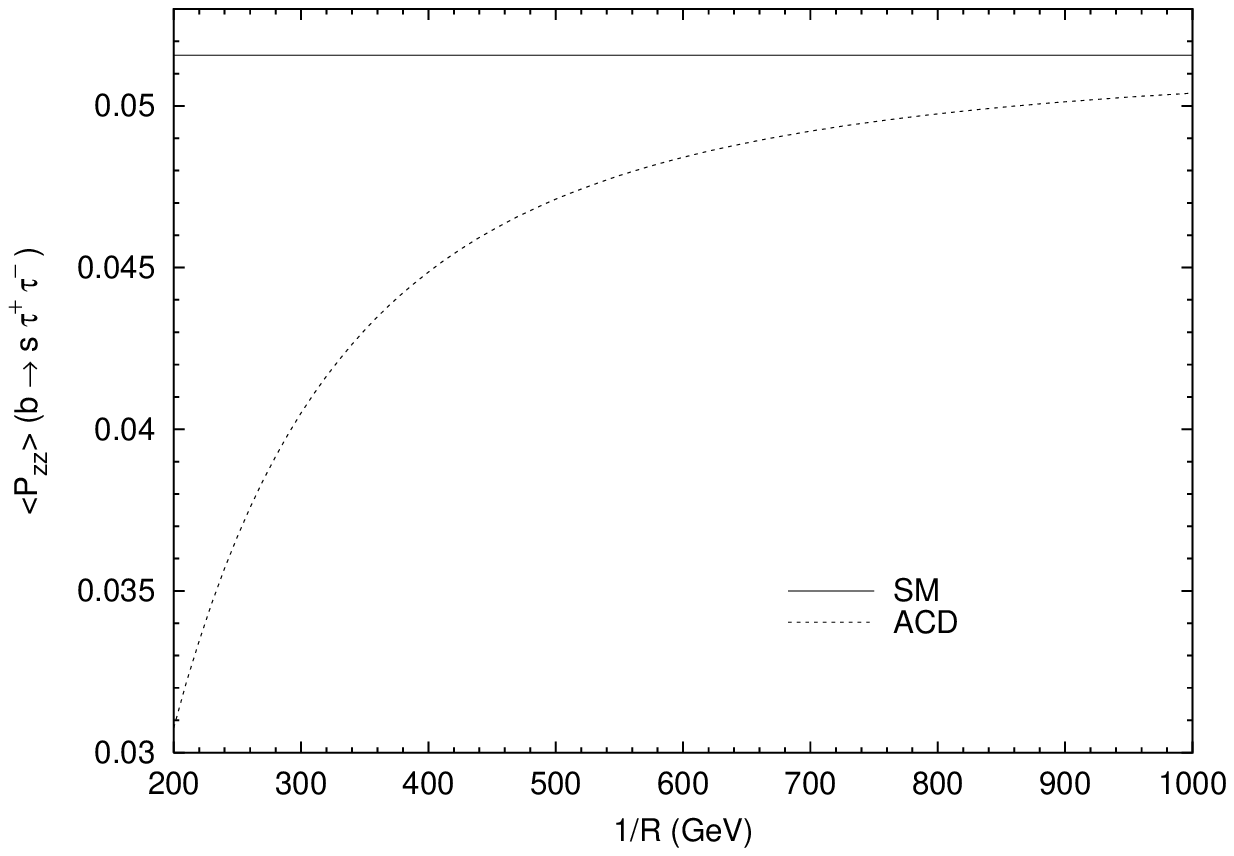}
\vskip 7.8 cm \caption{}
\end{figure}


\newpage

\newpage
\begin{thebibliography}{9}

\bibitem{Arkani}N. Arkani-Hamed, S. Dimopoulos and G. Dvali, Phys. Lett. {\bf B 429}, 263 (1998);
Phys. Rev. {\bf D 59}, 086004 (1999); I. Antoniadis, N.
Arkani-Hamed, S. Dimopoulos and G. Dvali, Phys. Lett. {\bf B 439},
257 (1998).
\bibitem{Colider}C. Macesanu, C.D. McMullen, S. Nandi, arXiv:hep-ph/0201300.
\bibitem{Antoniadis} I. Antoniadis, Phys. Lett. {\bf B 246} 377
(1990).
\bibitem{ACD} T. Appelquist, H. C. Cheng and B. A. Dobrescu, Phys. Rev.
\textbf{D 64}, 035002 (2001).

\bibitem{Colangelo:2006gv}
  P.~Colangelo, F.~De Fazio, R.~Ferrandes and T.~N.~Pham,
  Phys.\ Rev.\  {\bf D 74}, 115006 (2006)
  [arXiv:hep-ph/0610044].
\bibitem{Buras1} A.J. Buras, M. Spranger and A. Weiler, Nucl.\ Phys.\ {\bf B 660} (2003) 225.

\bibitem{Buras2} A.J. Buras, A. Poschenrieder, M. Spranger and A. Weiler, Nucl.\ Phys.\ {\bf
B 678} (2004) 455.
\bibitem{Haisch} U. Haisch and A. Weiler, Phys. Rev.
\textbf{D 76}, 034014 (2007).
\bibitem{Giri}R. Mohanata and A.K. Giri, Phys. Rev. {\bf D 75}, 035008, (2007).
\bibitem{Aliev} T.M. Aliev and M. Savci, Eur. Phys. J. {\bf C 50}, 91,(2007).
\bibitem{Pakistan} Ishtiaq Ahmed, M. Ali Paracha and M. Jamil Aslam, arXiv:0802.0740.
\bibitem{Hurth} T. Hurth,
 Rev. Mod. Phys. {\bf
75} (2003) 1159.

\bibitem{R4621} J. L. Hewett,
 Phys. Rev. {\bf D 53} (1996) 4964.
\bibitem{Aliev:2003fy}
  T.~M.~Aliev, V.~Bashiry and M.~Savci,
  Eur.\ Phys.\ J.\ {\bf C 35} (2004) 197.

  \bibitem{Aliev:2004hi}
  T.~M.~Aliev, V.~Bashiry and M.~Savci,
  JHEP {\bf 0405} (2004) 037
  [arXiv:hep-ph/0403282].

\bibitem{aali}A. Ali, Patricia Ball, L.T. Handoko and G. Hiller, Phys.Rev. {\bf D 61},074024 (2000).
\bibitem{R4622} F. Kr\"{u}ger and L. M. Sehgal
{\it Phys. Lett.} {\bf B 380} (1996) 199.
\bibitem{Bensalem:2002ni} W. Bensalem, D. London, N. Sinha and R. Sinha,
Phys. Rev. {\bf D 67} (2003) 034007.
\bibitem{Inami:1980fz}
T.~Inami and C.~S.~Lim,
Prog.\ Theor.\ Phys.\ {\bf 65}, 297 (1981) [Erratum-ibid.\ {\bf
65}, 1772 (1981)].
\bibitem{R23} A. J. Buras and M. M\"{u}nz,
{\it Phys. Rev.} {\bf D 52} (1995) 186.

\bibitem{R25} B. Grinstein, M. J. Savage and M. B. Wise,
{\it Nucl. Phys.} {\bf B 319} (1989) 271.

\bibitem{Willey} W. S. Hou, R. S. Willey and A. Soni, {\it Phys. Rev. Lett} {\bf
58} (1987) 1608; {\it ibid} {\bf 60} (1988) 2337 {\it Erratum}.

\bibitem{Deshpande}  N. G. Deshpande and J. Trampetic, {\it Phys. Rev. Lett} {\bf
60} (1988) 2583.
\bibitem{R5734} M. Jezabek and J. H. K\"{u}hn,
{\it Nucl. Phys.} {\bf B 320} (1989) 20.
\bibitem{misiak} M.\,Misiak, Nucl. Phys. {\bf B 393}\, (1993) 23.
\bibitem{misiakE}M.\,Misiak, Nucl. Phys. {\bf B 439}\, 461(E) \,(1995).
\bibitem{Bobeth:1999mk}
C.~Bobeth, M.~Misiak and J.~Urban,
Nucl.\ Phys.\ B {\bf 574}, 291 (2000).
\bibitem{Gambino:2003zm}
P.~Gambino, M.~Gorbahn and U.~Haisch,
Nucl.\ Phys.\ B {\bf 673}, 238 (2003).
\bibitem{NNLL}T. Huber, E. Lunghi, M. Misiak and D. Wyler, Nucl. Phys. {\bf
B 740} (2006) 105.
\bibitem{Fukae} S. Fukae and C.S. Kim, T. Yoshikawa, Phys. Rev. D {\bf 61} (2000) 074015.
\bibitem{R6016} G. Abbiendi {\it et. al}, OPAL Collaboration,
Phys. Lett. {\bf B 492}, 23 (2000).

\bibitem{R6017} A. Rouge,
Workshop on $\tau$ lepton physics, Orsay, France (1990).
\bibitem{Anirban} A. Kundu, arXiv:0806.3815.

\end{thebibliography}
\end{document}